\documentclass[aps,prl,twocolumn,superscriptaddress,showpacs]{revtex4-1}

\usepackage{amsmath,amssymb,graphics,epsfig,epstopdf,color,verbatim,ulem,braket,tabularx}
\usepackage[colorlinks,linkcolor=blue,citecolor=blue,urlcolor=blue]{hyperref}

\begin{document}

\title{Diagnosis of interaction-driven topological phase via exact diagonalization}

\author{Han-Qing Wu}
\affiliation{Department of Physics, Renmin University of China, Beijing 100872, China}
\author{Yuan-Yao He}
\affiliation{Department of Physics, Renmin University of China, Beijing 100872, China}
\author{Chen Fang}
\email{cfang@iphy.ac.cn}
\affiliation{Beijing National Laboratory for Condensed Matter Physics,
and Institute of Physics, Chinese Academy of Sciences, Beijing 100190, China}
\author{Zi Yang Meng}
\email{zymeng@iphy.ac.cn}
\affiliation{Beijing National Laboratory for Condensed Matter Physics,
and Institute of Physics, Chinese Academy of Sciences, Beijing 100190, China}
\author{Zhong-Yi Lu}
\affiliation{Department of Physics, Renmin University of China, Beijing 100872, China}

\begin{abstract}
We propose a general scheme for diagnosing interaction-driven topological phases in weak interaction regime using exact diagonalization (ED). The scheme comprises the analysis of eigenvalues of the point-group operators for the many-body eigenstates and the correlation functions for physical observables to extract the symmetries of the order parameters and the topological numbers of the underlying ground states at the thermodynamic limit from a relatively small size system afforded by ED. As a concrete example, we investigate the interaction effects on the half-filled spinless fermions on the checkerboard lattice with a quadratic band crossing point. Numerical results support the existence of a spontaneous quantum anomalous Hall phase purely driven by a nearest-neighbor weak repulsive interaction, separated from a nematic Mott insulator phase at strong repulsive interaction by a first-order phase transition.
\end{abstract}


\date{\today} \maketitle

\textit{Introduction} The pursuit of interaction-driven topological phases in fermions is becoming a collective activity in condensed matter physics community~\cite{Congjun2004,WenJun2010,Pesin2010,Castro2011,XuCenke2011,RanYing2011,Kurita2011,KSun2012,Venderbos2012,XuZhihao2013,
Dora2014,Herbut2014,Imada2014,Kitamura2015,YLWang2015,Maciejko2015,Kondo2015,Venderbos2015}, as people are expecting that such phases, if discovered, will combine both the richness of many-body effects and the elegance of topological physics. In Ref.~\onlinecite{Raghu2008}, Raghu and coworkers proposed the possibility of repulsive interaction generated current loops (spin-orbital coupling) in spinless (spin-1/2) electrons on a honeycomb lattice, which gives rise to quantum anomalous Hall (QAH) [quantum spin Hall (QSH)] phases. Although more recent analytical and numerical works~\cite{Weeks2010,Castro2013,Yongfei2013,Shimpei2014,Herbut2014,Hohenadler2014,Huaiming2014,Motruk2015,Lauchli2015,Scherer2015} have disputed the proposal in that particular model, alternative routes towards the realization of interaction-driven topological phases are currently being actively explored~\cite{ZhangFan2011,Andreas2011,Barnea2012,Kitamura2015,YLWang2015,Yoshida2014,Jiang2014,Jiang2015,Kurita2015,Venderbos2015,WZhu2015,Golor2015,Venderbos2016}.

\begin{figure}[htp!]
  \centering
  % insert JPEG file testjpg.jpg
  \includegraphics[width=0.9\columnwidth]{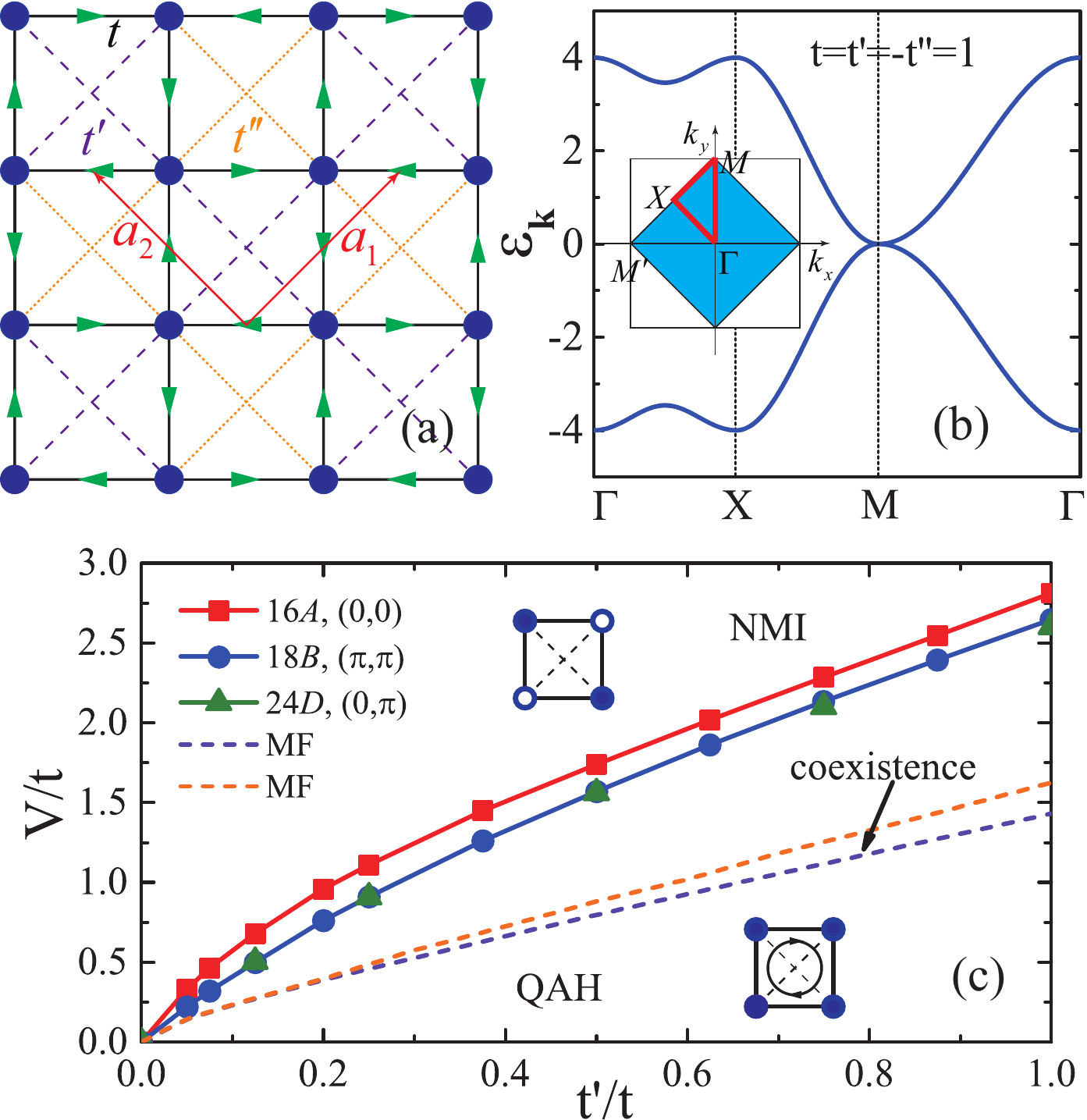}
  \caption{(Color online) (a). $16A$ cluster of checkerboard lattice. Two next-nearest-neighbour hopping amplitudes $t'$ and $t''$ are differentiated by the purple long-dashed and orange short-dashed lines. The green arrows represent the current loops in the spontaneous QAH phase. $\mathbf{a}_{1}=(1,1), \mathbf{a}_{2}=(-1,1)$ are the primitive vectors. (b). Noninteracting band structure along the high-symmetry path $\Gamma(0,0)\rightarrow X(\pi/2,\pi/2)\rightarrow M(0,\pi)\rightarrow \Gamma(0,0)$. Unlike the massless Dirac points in honeycomb lattice, the QBCP gives rise to a finite DOS. (c). The ground state phase diagram obtained from ED calculations. We use the level crossing (avoided level crossing) in the $16A$ ($18B$, $24D$) cluster under the periodic (anti-periodic, (0,$\pi$) twisted phase) boundary condition to determine the phase boundary. Phase boudaries determined from self-consistent mean-field calculation are also presented with dashed lines. Insets are the caricatures of the QAH and NMI order parameters in real-space.}
  \label{fig:chblatt}
\end{figure}

On the other hand, in a 2D system, unlike the Dirac point, a quadratic band crossing point (QBCP) with finite density of states (DOS) at the Fermi energy is unstable for arbitrarily weak interactions, leading to the possibility of spontaneous breaking of rotational symmetry (nematic phase) or time-reversal invariance~\cite{WZheng2008,Chong2008,KSun2008,KSun2009,Stefan2011,Dora2014,Murray2014}. In Ref.~\onlinecite{KSun2009}, K. Sun and coworkers proposed that the short-range repulsive interaction in spinless fermions is marginally relevant in one-loop renormalization group, and the leading mean-field instability is towards a QAH insulator with broken time-reversal symmetry. At the noninteracting limit, the QBCP acquires a dynamic critical exponent $z=2$, which renders the effective dimension of the underlying 2D system 4, and hence the corresponding mean-field analysis is likely to be permitted by the Ginzburg criterion~\cite{Ginzburg1960}.

To diagnose the interaction-driven topological phases, in this work, we design a scheme that enables us to extract definitive information on the thermodynamic ground state, including the symmetries of the phases and their topological numbers, from relatively small size systems studied by ED. Such a  diagnosis scheme is comprised of the analyses of eigenvalues of the point-group operators for the many-body eigenstates and the correlation functions for physical observables. We apply this scheme to the half-filled spinless fermions on the checkerboard lattice with a quadratic band crossing point.~\footnote{The possible interaction-driven topological phases have also been studies on other 2D lattices with QBCP, like kagome lattice~\cite{WenJun2010, LiuQin2010}, Lieb lattice~\cite{WFTsai2015,Dauphin2015} and honeycomb lattice~\cite{Venderbos2016,Kurita2015}.} We map out the full phase diagram in the parameter space with two gapped phases: a time-reversal breaking QAH phase at small repulsive interaction and a rotation symmetry breaking site-nematic Mott insulator (NMI) phase at large repulsive interaction, which are separated by a first-order quantum phase transition. This is the first time that the eigenvalues of the many-body eigenstates are used to infer the topological numbers in ED, and we remark that similar method can be used to diagnose other topological phases in weak interaction regime, such as the quantum spin Hall state and the $p+ip$ superconducting state.

\textit{Model and Method.} The system studied in this paper has the following Hamiltonian,
\begin{equation}
  \hat{H}=-\sum_{ij}\left(t_{ij}\hat{c}_{i}^{\dagger}\hat{c}_{j}+H.c.\right)+\mu\sum_{i}\hat{n}_{i}+V\sum_{<ij>}\hat{n}_{i}\hat{n}_{j}
  \label{eq:ChbLatt}
\end{equation}
where $t_{ij}$ is the hopping amplitude between sites $i$ and $j$, and $V$ is the nearest-neighbor repulsion. As shown in Fig.~\ref{fig:chblatt} (a), $t_{ij}=t,t',t''$ respectively, standing for the nearest ($t$, black solid lines), one type of next-nearest ($t'$, purple long-dashed lines) and the other type of next-nearest ($t''$, yellow short-dashed lines) neighbor hopping amplitudes. We set $t'=-t''$ to achieve the particle-hole symmetry (although our results also hold for the non-particle-hole symmetric case) and set chemical potential $\mu=-2V$ to guarantee half-filling~\cite{KSun2008,KSun2009}. To simplify the notation, the nearest-neighbor hopping $t$ and the nearest-neigbhor bond length $a$ are set to be units of energy and length.

The model in Eq.~\ref{eq:ChbLatt} accquires $C_{4}$ point-group symmetry and time-reversal symmetry $T$. The QBCP at $M$ point (shown in Fig.~\ref{fig:chblatt} (b)) with monopole flux $2\pi$ in the noninteracting band structure is protected by the combined symmetry of $T$ and $C_{4}$~\cite{Chong2008, KSun2008, KSun2009, JMHou2013}. In the ED calculations, we employed clusters with four different geometries (denoted as $16A$, $18B$, $24C$ and $24D$, as shown in the Section I of Supplemental Material~\cite{Suppl}). The results in the main text, especially the analysis of eigenvalues of the $C_4$ operators, are mainly obtained from the $16A$ cluster which respects the full symmetries of the Hamiltonian. Some physical observables of other clusters, particularly the $18B$ cluster which is also respects the $C_{4}$ symmetry, are also presented. For a given cluster, we apply a chosen set of twisted phases at the boundaries to ensure that the QBCP is included at the discretized single-particle momenta. Since the QBCP is the Fermi surface at half-filling, for small size calculations, it is crucial to include the states on the Fermi surface. Supplemental Material~\cite{Suppl} explain in detail on the choice of the twisted phases.

\textit{Numerical results.} Our ED calculations provide the energies of the low lying eigenstates in the parameter space spanned by $V/t$ and $t'/t$, from which two gapped phases are identified. In each phase, the two lowest lying states are separated from the higher states by a spectral gap.
The two lowest lying states in each phase are thus identified as its the ground state subspace, from which the symmetry-breaking ground state arises in the thermodynamic limit. As will become clear later, the two gapped phases are distinct as their ground state subspaces have different representations of the $C_4$-symmetry.

\begin{figure}[htp!]
  \centering
  %% insert JPEG file testjpg.jpg
  \includegraphics[width=0.8\columnwidth]{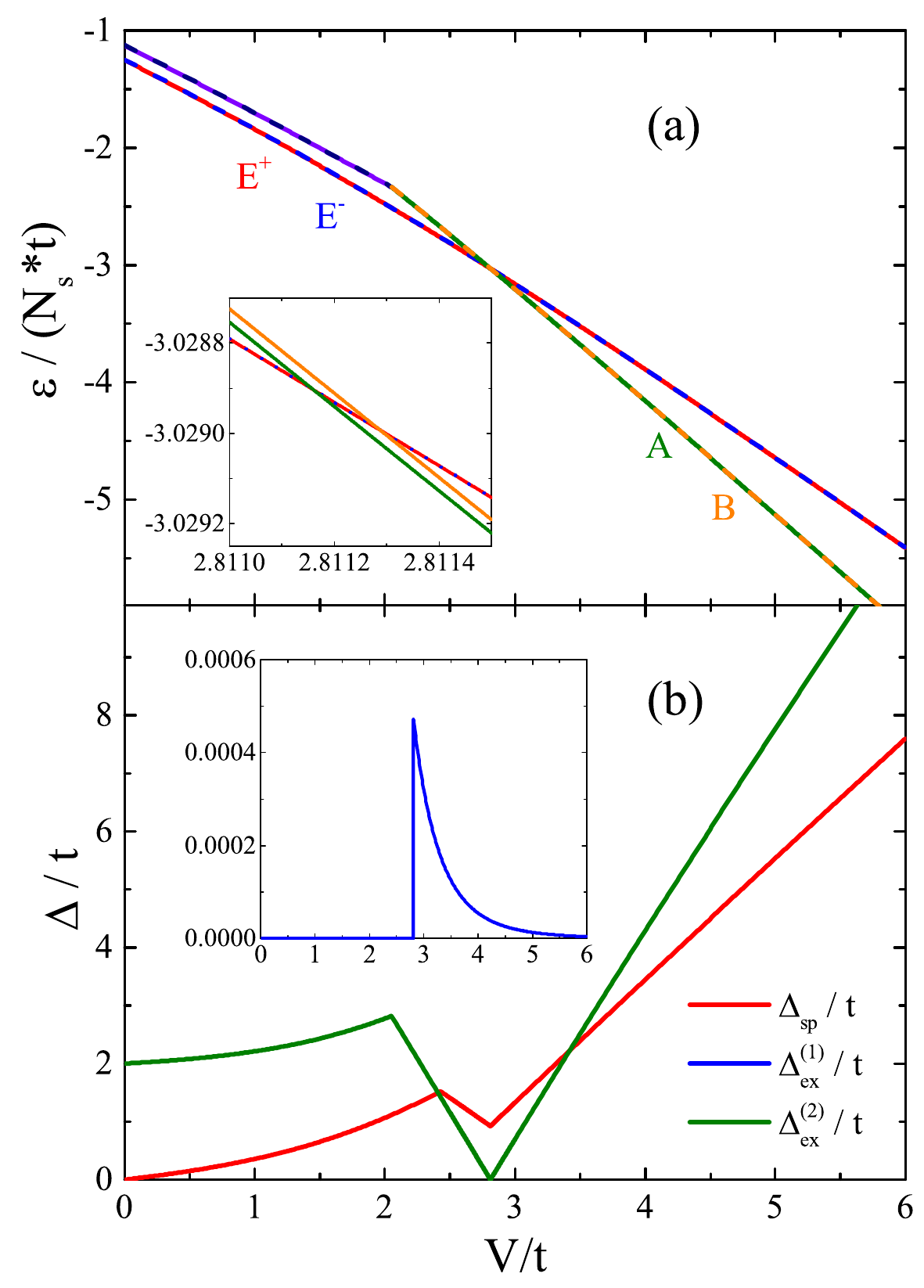}
  \caption{(Color online) (a) Energy density of the four lowest energy levels in the $16A$ as a function of $V/t$. The inset is a zoomin of the level crossing close to $V_{c}\approx 2.81t$. $E^{\pm}$, $A$ and $B$ label the four energy levels with their corresponding $C_{4}$ representation. (b) Single-particle gap $\Delta_{\text{sp}}=\left(E_{0}^{N_{e}+1}+E_{0}^{N_{e}-1}-2E_{0}^{N_{e}}\right)/2$ and the excitation gaps $\Delta_{\text{ex}}^{(n)}=E_{n}-E_{0}$ as a function of $V/t$. The single-particle gap opens at infinitesimal $V/t$ and has dip at $V_{c}/t$. While $\Delta_{ex}^{(2)}$ closes and reopens at $V_{c}/t$. Inset shows $\Delta_{ex}^{(1)}$ is exactly 0 (indicating $E^{\pm}$ are exactly degenerate) at $V<V_c$ and is actually finite (indicating $A$ and $B$ are only quasi-degenerate) at $V>V_c$.}
  \label{fig:EDGap}
\end{figure}

We first present the results for $t'/t=1$ for concision. Fig.~\ref{fig:EDGap} (a) shows the low energy spectra as a function of $V/t$. At small $V/t$, we can see an exact twofold ground state degeneracy, and these two degenerate ground states form the basis of the 2D $E^{\pm}$-representation of $C_{4}$ point group. This property is actually inherited from the Slater determinant state in the noninteracting limit. It can be explicitly checked that in the noninteracting limit, the Slater determinants on a finite square lattice with periodic boundary condition (PBC) form $E^\pm$ of $C_4$; as interaction turns on, the doublet is gapped from the higher states by a finite gap, thus remaining as the same 2D representation. At $V=V_{c}\approx 2.81t$, a level crossing occurs, after which two nearly degenerate excited states become the lowest eigenstates accompanied by the closing and reopening of excitation gap $\Delta_{ex}^{(2)}$, as shown in Fig.~\ref{fig:EDGap} (b). At $V>V_c$, the two states in the ground state subspace belong to the 1D $A$- and $B$-representation of $C_4$, respectively.

The same holds qualitatively for other values of $t'/t$: as $V$ is turned on, the system immediately enters one gapped phase, referred to as the small-$V$ phase, whose ground state subspace forms $E^\pm$-representation of $C_4$. Further increasing $V$, the system goes through a quantum phase transition and enters another gapped phase, or the large-$V$ phase, whose ground state subspace includes one $A$- and one $B$-representations (see Table S1 of Section VII in Supplemental Material~\cite{Suppl}). The phase diagram is plotted in Fig.~\ref{fig:chblatt} (c), where the phase boundary is defined on where the representation of the ground state sector changes ($16A$) or the avoided level crossing happens ($18B, 24D$). Having the phase boundaries determined, from here on, we employ our diagnosis scheme to answer the more physical questions: (i) what is the symmetry of the thermodynamic ground state? and (ii) what is the topological number, if any, of the ground state?

We first examine the small-$V$ phase and focus on the 16A cluster, whose ground state sector has two states with $C_4$ eigenvalues $+i$ and $-i$, denoted by $E^{\pm}$. These two states are exactly degenerate due to time-reversal symmetry, because $T$ sends a $C_4$-eigenstate of eigenvalue $+i$ to another one of eigenvalue $-i$. The symmetries of the Hamiltonian, $T$ and $C_4$, may either be preserved or broken in the thermodynamic limit: case-(a) the ground state is an eigenstate of $C_4$, thus breaking $T$ and preserving $C_4$; case-(b) the ground state is an equal weight superposition of $E^{\pm}$, thus breaking $C_4$ down to $C_2$, and as the two states have the same $C_2$ eigenvalue $C_2=C_4^2=(\pm{i})^2=-1$, suggesting a nematic phase.

Now we show that only case-(a) is possible for the small-$V$ phase and one can never have a thermodynamic ground state that is a superposition of the $C_4$ eigenstates with eigenvalues $\pm{i}$. To see this, we first calculate the Chern numbers of the $C_4$-eigenstates. For a finite system, the Chern number may be defined via its linear response to a twisted phase at the boundaries~\cite{NiuQian1985,Fukui2005,Varney2011}. In Ref.~\onlinecite{CFang2012}, it was shown that, in a weakly interacting system, for any gapped state that is an eigenstate of some rotation operator, its Chern number is directly related to the rotation eigenvalue under periodic boundary condition without twisted phases. %which spares us the need to extract reliable information on the ground state subspace for generic twisted phases (see Supplemental Material~\cite{Suppl} for a discussion of the effect of the twisted phase).
Our numerical data suggest that the small-$V$ phase extends to $V=0$ and is a gapped phase with weak interaction, so the Chern number $C$ is determined by the $C_4$-eigenvalue $\xi=1,-1,i,-i$ up to a multiple of 4
\begin{equation}\label{eq:Chern}
i^C=\xi.
\end{equation}
Using this formula, we determine the Chern numbers of the two lowest lying states $E^{\pm}$ as (see Table \ref{tab:table1}).
\begin{table}%The best place to locate the table environment is directly after its first reference in text
\caption{\label{tab:table1}Symmetry properties of many-body eigenstates of $16A$ under $C_4$. SSB stands for spontaneously symmetry breaking.}
\begin{ruledtabular}
\begin{tabular}{cccc}
interaction  &$\xi(0,0)$    & SSB                         	     &Chern number\\
\colrule
$V<V_{c}$ & $\pm i$       & TRS                         	     & $\pm 1$\\
$V>V_{c}$ & $\pm1$       & $C_{4}\rightarrow{C}_2$     & 0 \\
\end{tabular}
\end{ruledtabular}
\end{table}
\begin{equation}
\begin{split}
C_{E^+}&=1\;\textrm{mod}\;4,\\
\nonumber
C_{E^-}&=-1\;\textrm{mod}\;4.
\end{split}
\end{equation}
Next, we argue that the small-$V$ thermodynamic limit ground state cannot be a superposition of $E^{\pm}$. Because if it were the case, since we have just shown $E^{\pm}$ have different Chern number, their superposition would imply that the thermodynamic limit ground state has ambiguous Chern number, but this is against the general principle that the ground state of any gapped system should carry unique Chern number (see Section II and III in Supplemental Material~\cite{Suppl} for detailed discussion). Therefore, the thermodynamic ground state can only be one of $E^{\pm}$ with a nonzero Chern number. The small-$V$ phase hence breaks $T$ and preserves $C_4$, and carries Chern number of $\pm1$ up to a multiple of 4. The small-$V$ phase is an interaction-induced QAH state.

For the large-$V$ phase, the two lowest lying states are quasi-degenerate: there is a small gap in between that scales with the size of the system to some inverse power. The two states have $C_4$-eigenvalues of $+1$ and $-1$ respectively, or belong to the 1D $A$- and $B$-representation of $C_4$. The formula Eq.(\ref{eq:Chern}) no longer applies in this phase due to the strong interaction. Fortunately, deep in this phase there is a large gap separating the two lowest states from the other part of the spectrum for arbitrary twisted phase (see Fig. 3S in Supplemental Material~\cite{Suppl}), therefore we can use the winding of the wavefunction under different twisted phase to calculate the Chern number, which turns out to be zero for both $A$- and $B$-states. Therefore, any $C_4$-breaking local operator may have off-block-diagonal elements in the lowest lying subspace. The thermodynamic ground state is hence a superposition of the two states, which breaks $C_4$ yet preserves $C_2$. Whether or not the thermodynamic ground state breaks $T$ depends on the relative phase in the coefficients of the superposition. Our ED calculation shows that the matrix elements of the bond current operator are extremely small for large $V/t$ (not shown). This additional evidence pins down the large-$V$ phase to an NMI, which has zero Chern number. In order for the Chern number changes by an odd integer, a topologically protected level crossing must occur at some special twisted-boundary condition when the system has space-inversion symmetry or higher~\cite{Varney2011}. That is why we can see a level crossing in the $16A$ cluster calculation under PBC, as shown in Fig.~\ref{fig:EDGap} (a).

The above analysis on the ED results help us extract information on the symmetry and the topology of the thermodynamic ground state. It does not, however, give the form of the leading order parameters and the corresponding electronic structures of the phases. To this end, we also perform a mean-field study following Ref.~\onlinecite{KSun2009}, which generates the mean-field phase boundaries in Fig.~\ref{fig:chblatt} (c). The leading order parameters for the two phases are the current loop and the site nematicity defined as
\begin{equation}
\begin{split}
m_{\text{QAH}}&=\frac{1}{4}\sum_{\delta=\pm\hat{x},\pm\hat{y}}D_\delta\langle\hat{J}_{i,i+\delta}\rangle,\\
m_{\text{NMI}}&=\frac{1}{4}\sum_{\delta=\pm\hat{x},\pm\hat{y}}\langle\hat{\rho}_{i,i+\delta}\rangle,
\end{split}
\end{equation}
where $i$ labels the sites in the A-sublattice and $D_\delta=+1$ for $\delta=\pm\hat{x}$ and $-1$ for $\delta=\pm\hat{y}$. $\hat{J}_{i,i+\delta}=i(\hat{c}_{i}^{\dagger}\hat{c}_{i+\delta}-\hat{c}_{i+\delta}^{\dagger}\hat{c}_{i})$ is the current operator. $\hat{\rho}_{i,i+\delta}=\hat{c}^\dag_i\hat{c}_i-\hat{c}^\dag_{i+\delta}\hat{c}_{i+\delta}$ is the electron density difference between the A- and the B-sublattices. The caricatures of the ordered pattern are shown in the insets of Fig.~\ref{eq:ChbLatt} (c).

\begin{figure}[htp!]
  \centering
  %% insert JPEG file testjpg.jpg
  \includegraphics[width=0.8\columnwidth]{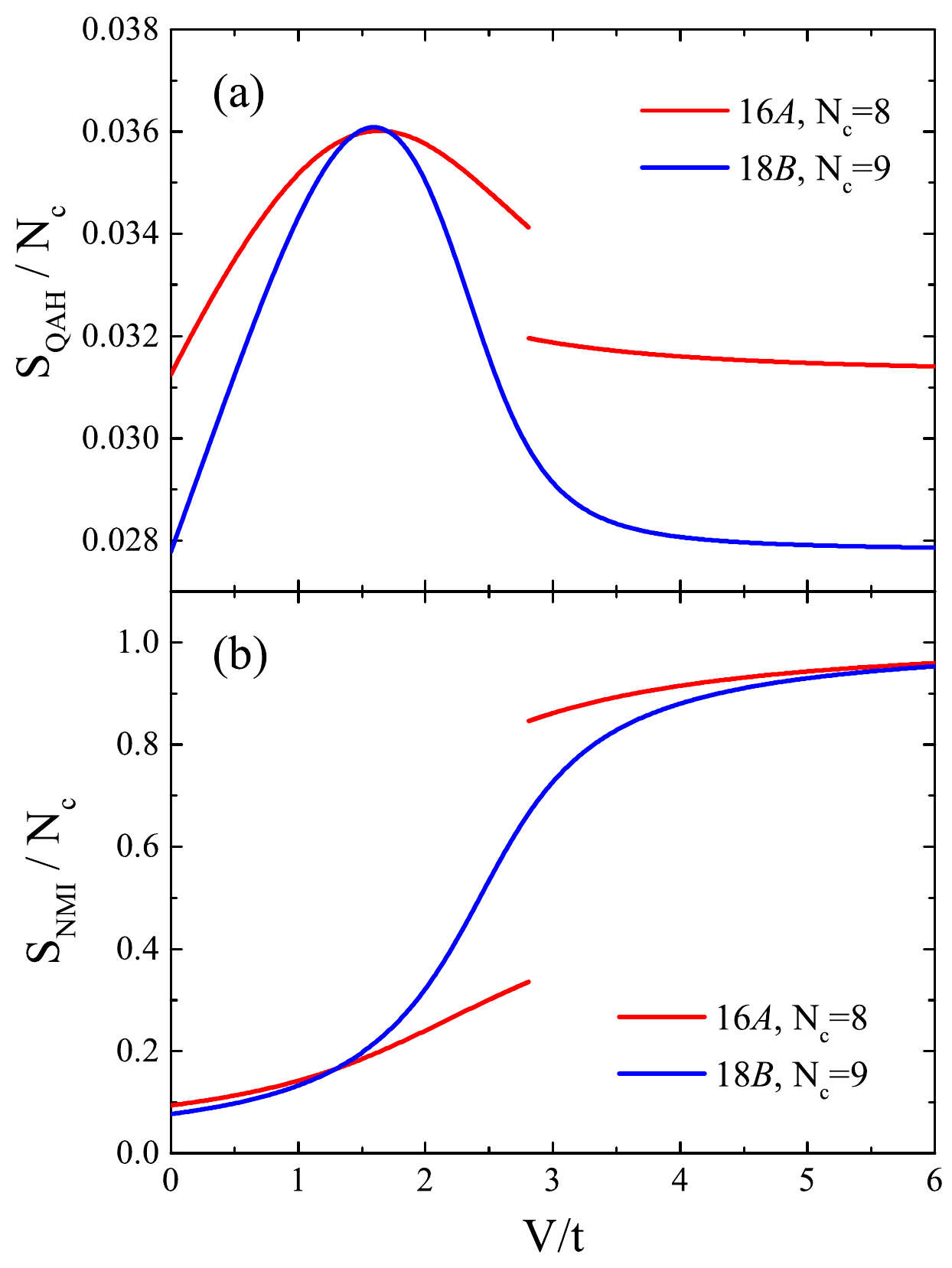}
  \caption{(Color online) The structure factors of (a) QAH and (b) NMI phases. The discontinuity in the $16A$ results is due to the level crossing. We clear see the enhancement of the QAH structure factor at $V<V_c$ and the saturation of the NMI structure factor at $V>V_c$.}
  \label{fig:EDStrFct}
\end{figure}

The mean-field phase diagram is qualitatively consistent with the ED results. However, it is fail to predict the insulating behavior of NMI when the site-nematic order parameter is small (see Sec.VI in Supplemental Material~\cite{Suppl}). Also we note that the ED results show a larger area of QAH phase, indicating an overestimate of the site-nematic order in the mean-field calculations. More importantly, we also computed the correlation functions of the order parameters in ED
\begin{eqnarray}
  S_{\text{QAH}} &=&\frac{1}{4}\sum_{i\in A}\sum_{\delta}D_{\delta}\langle\hat{J}_{i,i+\delta}\hat{J}_{i_{0},i_{0}+\delta_{0}}\rangle,\nonumber\\
  S_{\text{NMI}} &=& \frac{1}{4}\sum_{i\in A}\sum_{\delta}\langle\hat{\rho}_{i,i+\delta}\hat{\rho}_{i_{0},i_{0}+\delta_{0}}\rangle,
\end{eqnarray}
where we have used the translation symmetry and $i_{0},i_{0}+\delta_{0}$ is the reference bond. For comparison, here we present the results along the line $t=t'$ and plot the correlation functions versus $V/t$ in Fig.~\ref{fig:EDStrFct} . At small $V/t$, the broad peak in the QAH current loop structure factor (see Fig.~\ref{fig:EDStrFct} (a)) signifies that the QAH phase will be stable in the thermodynamic limit. The possibility of a bond-nematic phase in the small $V/t$ is also considered, but its correlation is clearly short-ranged (see Section V in Supplemental material~\cite{Suppl}). At large $V/t$, $S_{\text{NMI}}$ quickly increases and it saturates at $S_\textrm{NMI}=1$ in the $V/t\rightarrow\infty$ (see Fig.~\ref{fig:EDStrFct}(b)), indicating that all electrons are located at either A-sites or B-sites.

\textit{Discussion} Finally, we discuss the transition between the small-$V$ QAH and the large-$V$ NMI phases. The QAH phase preserves $C_4$ and breaks $T$, while the NMI phase breaks $C_4$ and preserves $T$. Therefore, they can either be separated by a first-order transition line or a region of coexisting phase (breaking both T and $C_4$). In the mean-field calculation (see Fig.~\ref{fig:chblatt} or in Ref.~\cite{KSun2009}), there is a very small region where both order parameters are non-vanishing, while the data from ED is insufficient to draw any conclusion. We conjecture there is a first-order transition. If there were a coexisting phase, the thermodynamic ground state presumably arises from the joint ground state subspaces of QAH and NMI, i.e., a linear superposition of the $E^\pm$-representation and the $A$- and $B$-representations of $C_4$. But we know that due to the difference in Chern numbers, only the $A$- and the $B$-representations, both having vanishing Chern number, can be linearly superimposed. In other words, the thermodynamic ground state cannot have a finite Chern number while being a superposition of different representations of $C_4$, so the QAH phase must preserve $C_4$ and cannot coexist with the NMI phase.

\begin{acknowledgments}
We would like to thank Zheng-Xin Liu and Lei Wang for helpful discussion. H.Q.W. would like to thank Rong-Qiang He for helpful discussions about the Lanczos exact diagonalization method. Z.Y.M. is particularly indebted to D. Wang and H. Fan for their early engagement and stimulating discussions that motivate this study. The numerical calculations were carried out at the Physical Laboratory of High Performance Computing in RUC as well as the National Supercomputer Center in Guangzhou on the Tianhe-2 platform.  C. F. and Z. Y. M. acknowledge support from the Ministry of Science and Technology (MOST) of China under Grants No. 2016YFA0302400 and No. 2016YFA0300502. H. Q. W., Y. Y. H., Z. Y. M., and Z. Y. L. acknowledge support from the National Natural Science  Foundation  of  China  (NSFC)  under  Grants No.  11474356,  No.  91421304,  No.  11421092,  and No. 11574359. C. F. and Z. Y. M. are also supported by the National Thousand-Young-Talents Program of China.

\end{acknowledgments}

{\it Note added.} We recently become aware of an interesting work~\cite{WZhu2016} where interaction-driven spontaneous quantum Hall effect is obversed on kagome lattice via ED and DMRG.

\bibliography{QAHBib}

\clearpage

%\widetext
\begin{center}
\textbf{\large Supplemental Material: Diagnosis of interaction-driven topological phase via exact diagonalization}
\end{center}
%%%%%%%%%% Merge with supplemental materials %%%%%%%%%%
%%%%%%%%%% Prefix a "S" to all equations, figures, tables and reset the counter %%%%%%%%%%
\setcounter{equation}{0}
\setcounter{figure}{0}
\setcounter{table}{0}
\setcounter{page}{1}
\makeatletter
\renewcommand{\theequation}{S\arabic{equation}}
\renewcommand{\thefigure}{S\arabic{figure}}
\renewcommand{\thetable}{S\arabic{table}}
\renewcommand{\bibnumfmt}[1]{[S#1]}
\renewcommand{\citenumfont}[1]{S#1}
%%%%%%%%%% Prefix a "S" to all equations, figures, tables and reset the counter %%%%%%%%%%

\section{I. Finite-size clusters used in the ED calculations}
\begin{figure}[h]
  \centering
  %% insert JPEG file testjpg.jpg
  \includegraphics[width=1.0\columnwidth]{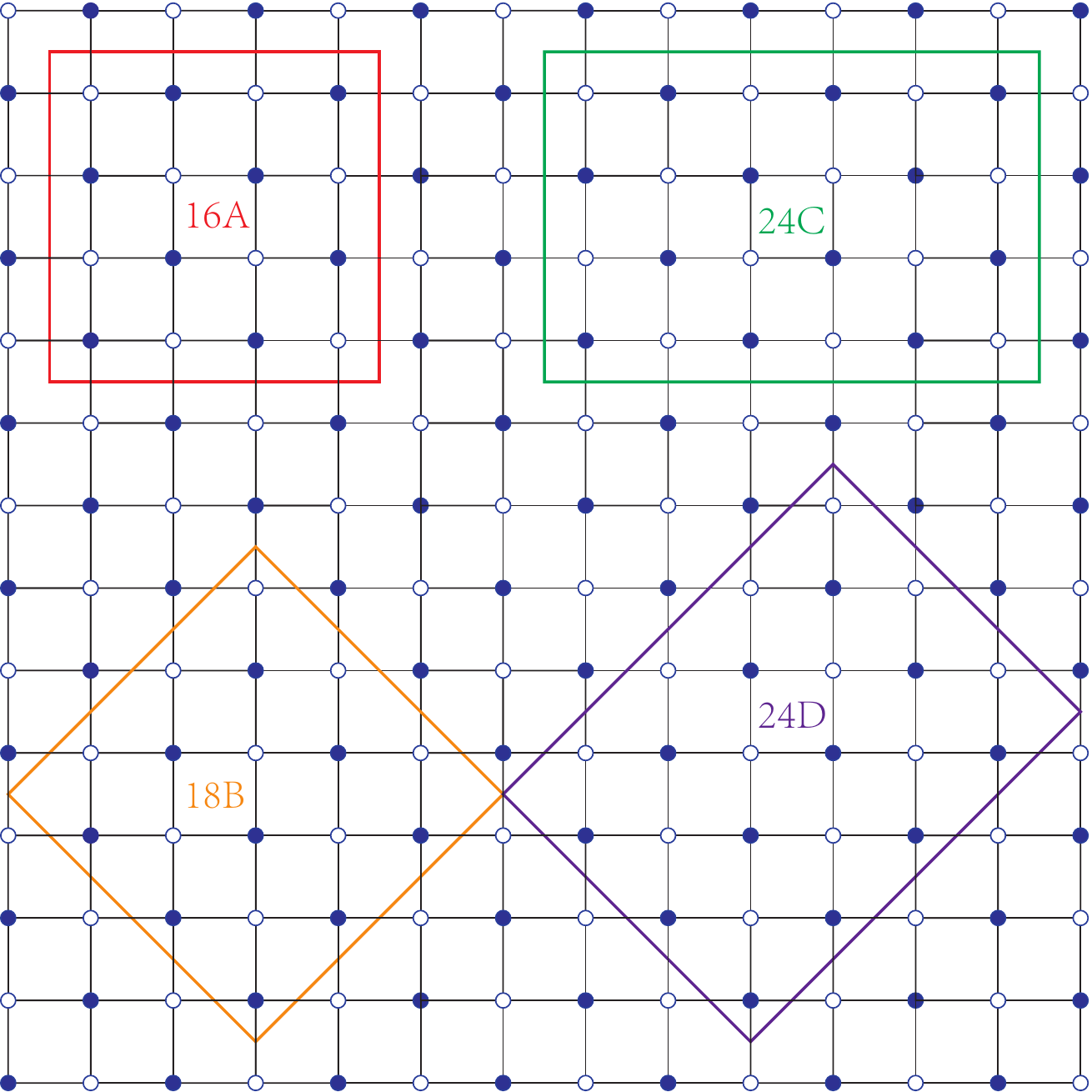}
  \caption{(color online) Four clusters used in our ED calculations. We use the full and empty circles to distinguish two sublattices within an unit cell.}
  \label{fig:clusters}
\end{figure}

Four clusters are used in our ED calculations which are shown in Fig.~\ref{fig:clusters}, denoted as $16A$, $18B$, $24C$ and $24D$, respectively. The $16A$ and $18B$ clusters respect the $C_{4}$ rotation symmetry, while other clusters, like $24C$ and $24D$, are not. The $16A$ and $24C$ clusters can capture the QBCP under the periodic boundary condition (PBC). $18B$ cluster captures the QBCP only in the anti-periodic boundary condition (anti-PBC) or twisted phase boundary condition with ($\pi$,$\pi$) twisted phase. $24D$ cluster in the (0,$\pi$) twisted phase boundary condition can also capture the QBCP.

\begin{figure}[h]
  \centering
  %% insert JPEG file testjpg.jpg
  \includegraphics[width=1.0\columnwidth]{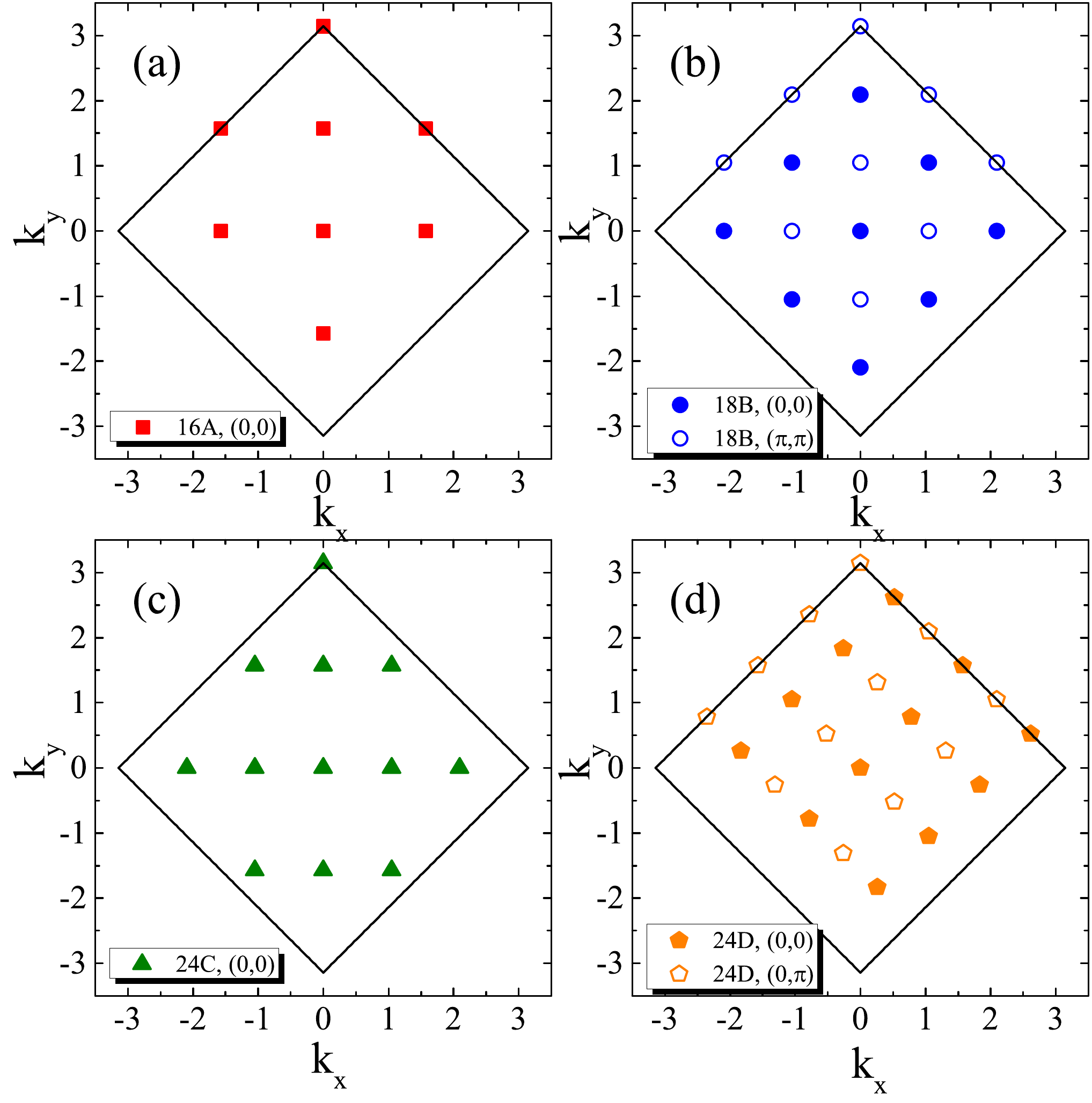}
  \caption{(color online) Illustration of the $\mathbf{k}$ mesh in the Brillouin zone (BZ) of finite-size clusters. The solid (hollow) points represent the effective $\mathbf{k}$ points under periodic (twisted phase) boundary condition [with twisted phase $(\phi_{x},\phi_{y})$].}
  \label{fig:kpoints}
\end{figure}

\section{II. The effect of twisted phase at the boundary}

In calculation on small-size clusters, it is important to include the single-particle states that are at or close to the Fermi surface. However, for clusters $18A$, $24C$ and $24D$ studied in this work, the periodic boundary condition results in a set of $k$-points that does not contain $M=(0,\pi)$ which is the Fermi surface for the half-filling. To remedy this, we introduce twisted phases at the boundary of the system, such as any hopping that crosses the $x$-boundary ($y$-boundary) gains an additional phase of $\phi_x$ ($\phi_y$). This additional phase moves the $k$-points, and for a certain choice of $(\phi_x,\phi_y)$, one is able to include $M$-point in the set of allowed $k$-points. In Fig.~\ref{fig:kpoints}, we plot the allowed momenta for the four clusters at $(\phi_x,\phi_y)=(0,0),(\pi,\pi),(0,0),(\pi,0)$ respectively. We see that by these choices we can always capture the $M$-point in the calculation.

\begin{figure}
  \centering
  %% insert JPEG file testjpg.jpg
  \includegraphics[width=\columnwidth]{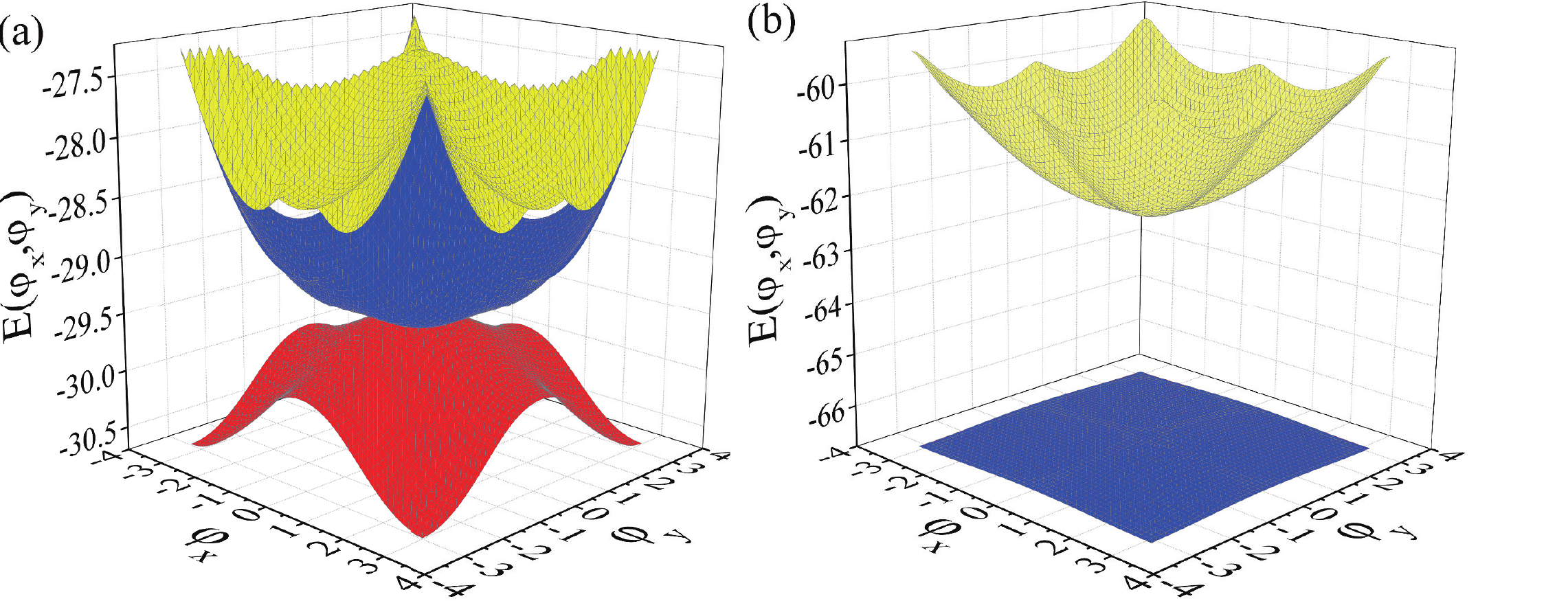}
  \caption{(color online) The energy of the three lowest states in 16A cluster as a function of $(\phi_x,\phi_y)$ at (a) $V=t$ and (b) $V=4t$, respectively.}
  \label{fig:TwistEnergy}
\end{figure}

The choice of the twisted phase is crucial for the small-$V$ phase, where interaction is weak and the electrons can still be considered as Bloch electrons. It turns out that the system is sensitive to the twisted phases at the boundary and the low energy spectra strongly depends on $(\phi_x,\phi_y)$. In Fig.~\ref{fig:TwistEnergy} (a), we plotted, for the 16A cluster, the energies of the three lowest states as functions of $(\phi_x,\phi_y)$ for $\phi_{x,y}\in(-\pi,\pi]$. In the thermodynamic limit, the allowed $k$-points form a continuum BZ, independent of the choice of twisted phases, so that the energy spectrum should be independent of $(\phi_x,\phi_y)$. We hence conclude that it takes a much larger cluster to capture the qualitative ground state information for all $(\phi_x,\phi_y)$. The Chern number of a many-body state $|\Omega\rangle$ can be calculated~\cite{NiuQian1985,Fukui2005,Varney2011} via
\begin{equation}\label{eq:winding}
\begin{split}
C=&\frac{i}{2\pi}\int{d}\phi_xd\phi_y\big[\frac{\partial}{\partial_{\phi_x}}\langle\Omega(\phi_x,\phi_y)|\frac{\partial}{\partial_{\phi_y}}|\Omega(\phi_x,\phi_y)\rangle \\
&-\frac{\partial}{\partial_{\phi_y}}\langle\Omega(\phi_x,\phi_y)|\frac{\partial}{\partial_{\phi_x}}|\Omega(\phi_x,\phi_y)\rangle\big].
\end{split}
\end{equation}
However, in our case, we cannot use Eq.(\ref{eq:winding}) to find the Chern numbers in the small-$V$ phase, because we cannot obtain the ground state manifold for arbitrary $\phi_{x,y}$ due to the small system size.

For the large-$V$ phase, Fig.~\ref{fig:TwistEnergy} (b) shows that the low energy spectra is significantly less dependent on the twisted phases at the boundary, and the two lowest states are always separated from the above by a large gap. We conjecture that for this highly localized phase (the correlation function of the site-nematic order parameter being almost saturated), the system is insensitive to the change of hopping at the boundary. We hence apply Eq.(\ref{eq:winding}) to calculate the Chern numbers of the two states in the ground state subspace, finding both to be zero.

\section{III. Thermodynamic ground state cannot be a superposition of two eigenstates having different topological numbers}

In this section, we attempt to provide a more formal argument for the statement that the thermodynamic ground state cannot be a linear superposition of two quantum eigenstates from the ground state subspace that have different topological numbers, e.g., Chern numbers in our case.

Generally, as all symmetries are explicitly preserved in ED, one cannot obtain the symmetry breaking thermodynamic ground state as the lowest lying state in the spectrum. Instead, one obtains a set of low lying states separated by a spectral gap from all elementary excitations, called the Anderson's tower of states. These states form the ground state subspace. The states within the subspace have energy separations that decay to zero faster than any Goldstone mode as the system size increases, and are hence called quasi-degenerate. In reality, in a thermodynamic system, a local random symmetry breaking field, although arbitrarily small, will hybridize the eigenstates in the subspace and lift the degeneracy, favoring a superposition state that breaks the symmetry as the true ground state.

In this process of symmetry breaking, it is necessary that a local random field can hybridize different eigenstates, that is
\begin{equation}
\langle{\Omega_i}|\hat{O}|\Omega_j\rangle\neq0,
\end{equation}
for some $i\neq{j}$, where $|\Omega_i\rangle$ is an eigenstate from the subspace and $\hat{O}$ an arbitrary local operator. Now we argue that if $|\Omega_i\rangle$ and $|\Omega_j\rangle$ have different topological numbers, the matrix element between $|\Omega_i\rangle$ and $|\Omega_j\rangle$ must vanish. Suppose it does not vanish, then starting from $|\Omega_i\rangle$, one can apply $\hat{O}$ for a finite duration, to rotate the state to $|\Omega_j\rangle$. This would imply that one can locally change the topological number of a large system, which is against the principle that one cannot change a topological quantum number locally. Therefore, as long as $|\Omega_{i,j}\rangle$ have different topological numbers, e.g., different Chern numbers, a local operator has vanishing matrix elements between the two states.

We conclude that, if the ground state subspace can be divided into topological sectors carrying different topological numbers, the thermodynamic ground state can only be a superposition state from a single topological sector.

\begin{figure}
  \centering
  %% insert JPEG file testjpg.jpg
  \includegraphics[width=0.8\columnwidth]{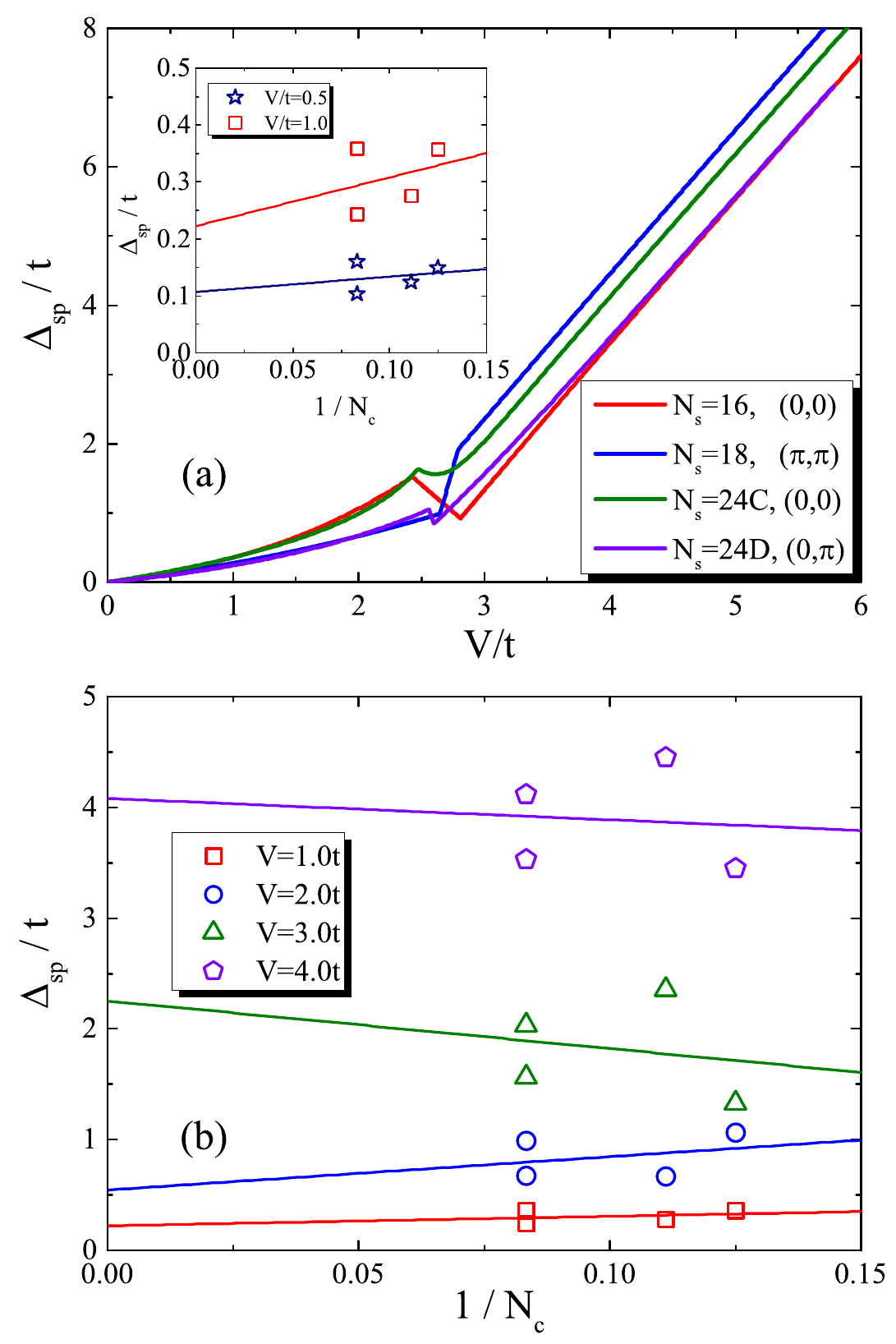}
  \caption{(color online) (a). The single-particle gap $\Delta_{\text{sp}}$ as a function of interaction strength $V/t$ for various finite-size clusters. $16A$ and $18B$ clusters which respect the $C_{4}$ symmetry get the lowest single-particle gap under the (0,0)- and ($\pi,\pi$)-twisted phase boundary condition when $V/t$ is small. the inset shows the linear extrapolation of the single-particle gap at $V=0.5t,1.0t$. (b). Linear extrapolation of the single-particle gap. For all the $V$-values, the $\Delta_{\text{sp}}$ are finite in the thermodynamic limit.}
  \label{fig:GapExtra}
\end{figure}

\begin{figure}
  \centering
  %% insert JPEG file testjpg.jpg
  \includegraphics[width=1.0\columnwidth]{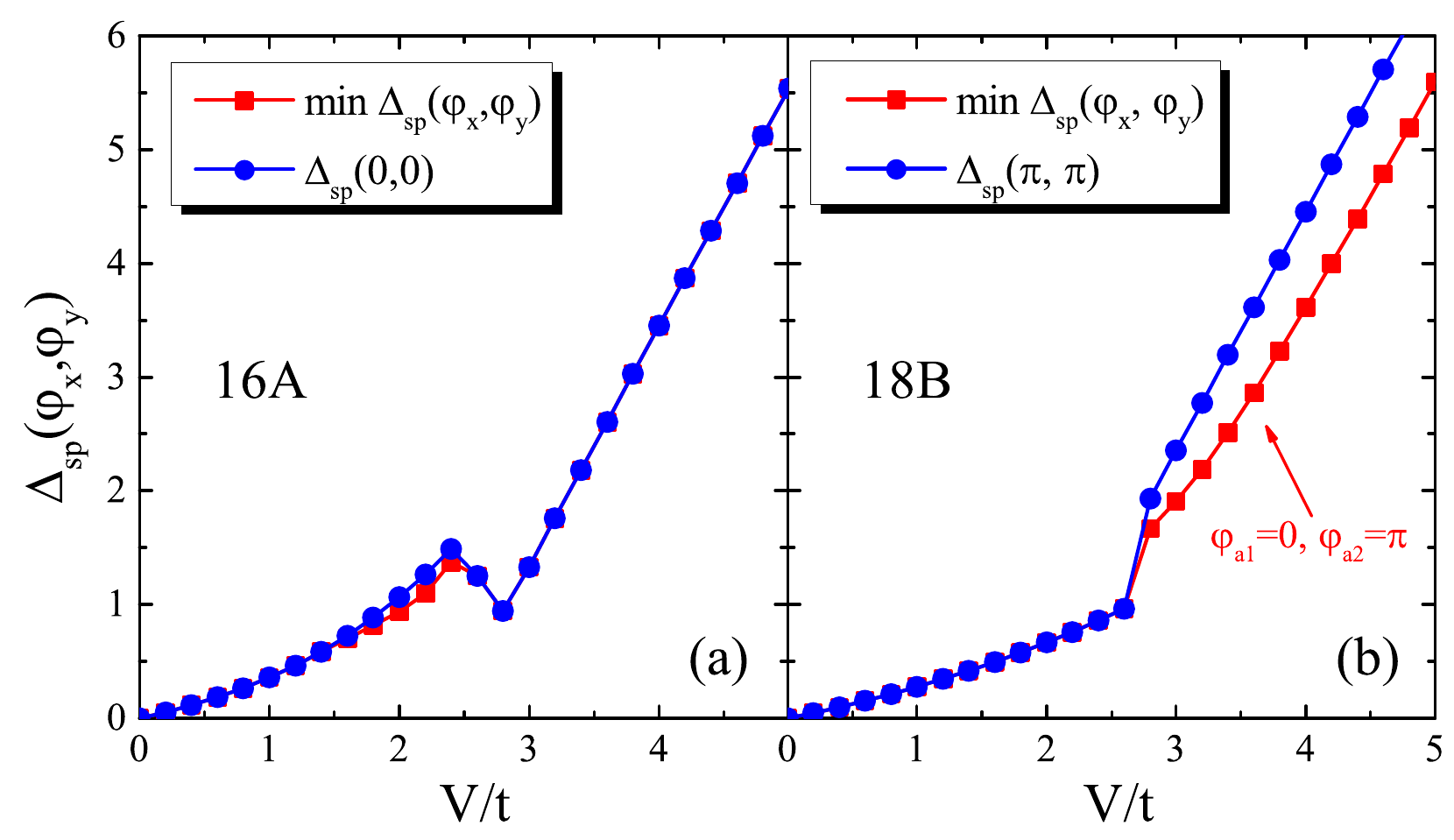}
  \caption{The minimum single-particle gap $\Delta_{\text{sp}}$ under different twisted phases for (a) $16A$ and (b) $18B$ clusters. In the small $V/t<1.5t$, the single-particle gap gets its minimum at QBCP (M Point) for both $16A$ and $18B$, while the single-particle gap gets its minimum at X point in the Brillouin zone at large $V/t$ NMI phase. These results are consistent with the mean-field results (see Fig.~\ref{fig:MFOrder}) for the small and large $V/t$ regime. However, an intermediate semi-metallic phase predicted by mean-field theory was not found in our ED calculations.}
  \label{fig:TwistPhase}
\end{figure}

\section{IV. Energy gap extrapolation}
In order to numerically confirm the insulating behaviors of the system at finite repulsive interaction, either in small-$V$ QAH or large-$V$ NMI phases. We perform an extrapolation of the single-particle gap with $1/N_{c}$, where $N_c$ is the number of unit cell contains in the finite ED clusters. As shown in Fig.~\ref{fig:GapExtra}, the single-particle gap $\Delta_{sp}$ scale to finite value all both small-$V$ and large-$V$.

To exclude the possible splitting of QBCP into two Dirac cones, we can use the twisted phase boundary condition to move the k-points. The numerical data in Fig.~\ref{fig:TwistPhase} show that, for the $16A$ ($18B$) cluster which respects the plaquette-centered C4 symmetry, the single-particle gap gets its minimum under the (0,0) [($\pi$, $\pi$)] twisted phase boundary condition in the small $V/t$ region. That indicates a direct opening of a single-particle gap at the QBCP. When the single-particle gap becomes large, where the single-particle gap gets its minimum in the Brillouin zone is physically irrelevant.

\section{V. Bond nematic structure factor}

\begin{figure}[h!]
  \centering
  %% insert JPEG file testjpg.jpg
  \includegraphics[width=0.9\columnwidth]{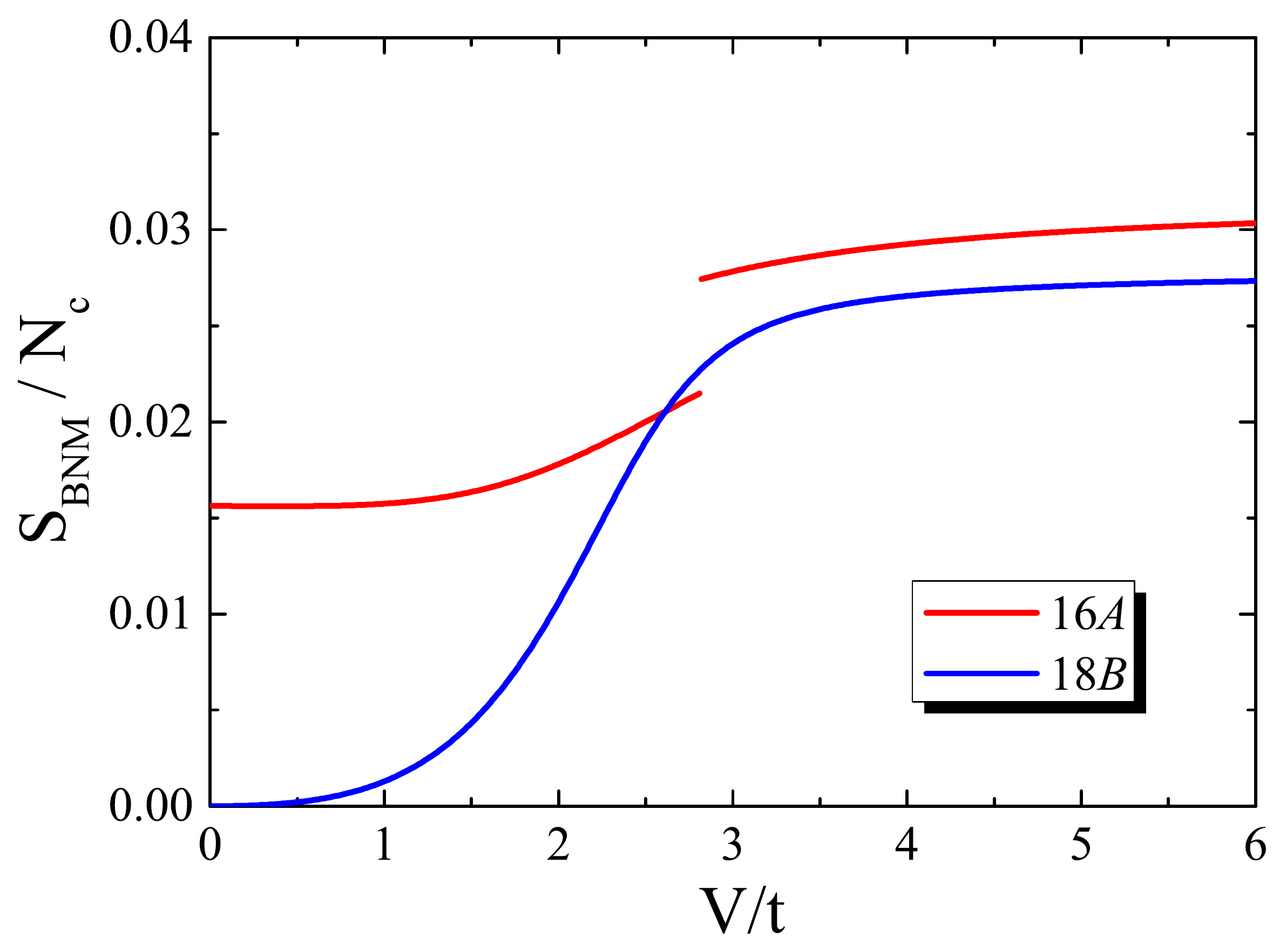}
  \caption{(color online) The bond nematic structure factor is shown for $16A, 18B$ clusters. $S_{BNM}\rightarrow1/4$ in the large $V/t$ is independent of cluster sizes. The bond-nematic phase correlation is clearly short-ranged in the parameter $0<V/t<V_{c}/t$ region.}
  \label{fig:BNMStrFct}
\end{figure}

The bond nematic (BNM) structure factor is defined as,
\begin{equation}
  S_{\text{BNM}} = \frac{1}{4}\sum_{i\in A}\sum_{\delta}D_{\delta}\braket{\mathcal{\hat{B}}_{i, i+\delta}\mathcal{\hat{B}}_{i_{0},i_{0}+\delta_{0}}}
\end{equation}
where $\mathcal{\hat{B}}_{i, i+\delta} = \hat{c}_{i}^{\dagger}\hat{c}_{i+\delta}+\hat{c}_{i+\delta}^{\dagger}\hat{c}_{i}$ is the bond operator between sublattice $i\in A$ and sublattice $i+\delta\in B$. $i_{0},i_{0}+\delta_{0}$ is the reference bond. $D_\delta=+1$ for $\delta=\pm\hat{x}$ and $-1$ for $\delta=\pm\hat{y}$. Fig.~\ref{fig:BNMStrFct} shows that at $V<V_c$, the BNM order parameter goes to zero as the cluster size increases.

\section{VI. Mean-field order parameters}

\begin{figure}[h]
  \centering
  %% insert JPEG file testjpg.jpg
  \includegraphics[width=0.8\columnwidth]{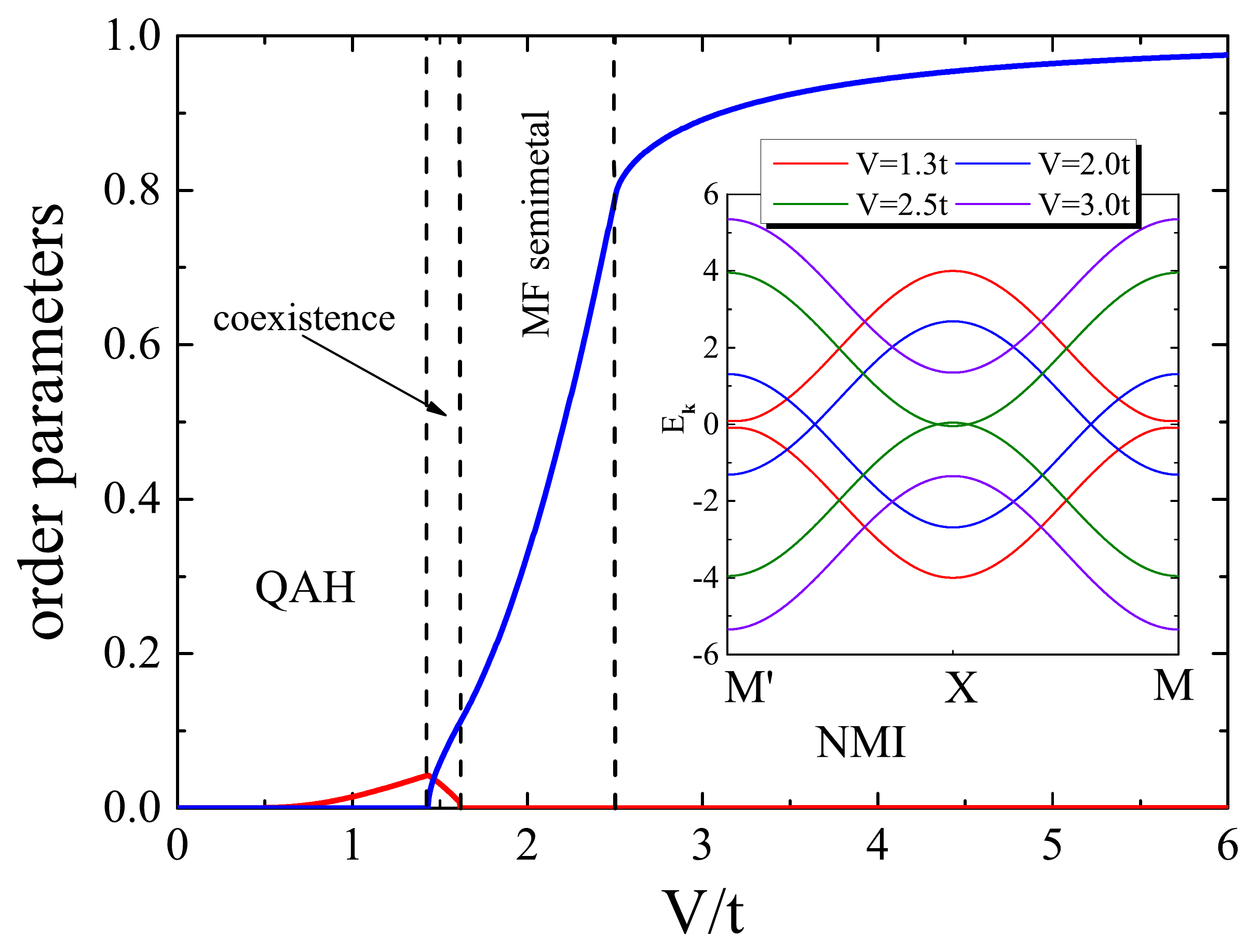}
  \caption{(color online) The mean-field results of QAH current loop (red line) and site-nematic (blue line) order parameters. We only present the results for $t'/t=1$ for concision. Besides the QAH and NMI, the mean-field results show an intermediate semi-metallic phase with two Dirac points along $M'(\pi,0)\rightarrow X(-\pi/2,\pi/2)\rightarrow M(0,\pi)$ path (see the inset).}
  \label{fig:MFOrder}
\end{figure}

In the mean-field calculations, we decouple the nearest-neighbor repulsive interaction term in the following way as
\begin{equation}
\begin{split}
  \hat{n}_{i}\hat{n}_{j} \mapsto &\braket{\hat{n}_{i}}\hat{n}_{j} + \braket{\hat{n}_{j}}\hat{n}_{i}-\braket{\hat{n}_{i}}\braket{\hat{n}_{j}}-\\
  &\braket{\hat{c}_{i}^{\dagger}\hat{c}_{j}}\hat{c}_{j}^{\dagger}\hat{c}_{i}-\braket{\hat{c}_{j}^{\dagger}\hat{c}_{i}}\hat{c}_{i}^{\dagger}\hat{c}_{j}+\left|\braket{\hat{c}_{i}^{\dagger}\hat{c}_{j}}\right|^{2}.
\end{split}
\end{equation}
This mean-field decoupled method can give rise to site-nematic, bond-nematic or QAH order. According to the mean-field and RG analysis in Ref.~\onlinecite{KSun2009}, our numerical analysis in the main text and the structure factor shown in Fig.~\ref{fig:BNMStrFct}, bond-nematic order is not favorable in the model we study. Therefore, we only consider the competition between the QAH current loop order and the site-nematic order which is shown in Fig.~\ref{fig:MFOrder}. Our numerical data combined with point-group symmetry analysis disagrees with the mean-field result that QAH and the site-nematic phase may coexist for a small parameter range.

\section{VII. four-fold rotation symmetry}

The $C_{4}$ point group is an Abelian group. And the generator is the clockwise $\pi/4$ rotation about the $z$ axis. The group character table is shown in Table~\ref{tab:table2}. As far as time-reversal symmetry is concerned, $E^{+}$ and $E^{-}$ form a two-dimensional irreducible representation.
\begin{table}[htp!]%The best place to locate the table environment is directly after its first reference in text
\caption{\label{tab:table2}The character table of $C_{4}$ point group.}
\begin{ruledtabular}
\begin{tabular}{ccccc}
$C_{4}$  &$E$    & $c_{2}$  & $c_{4}$      & $c_{4}^{3}$ \\
\colrule
$A$     & 1       & 1            & 1               & 1 \\
$B$     & 1       & 1            &-1               &-1 \\
$E^{+}$ & 1       &-1            & i               &-i \\
$E^{-}$ & 1       &-1            &-i               & i \\
\end{tabular}
\end{ruledtabular}
\end{table}

\end{document}